\documentclass[prd,twocolumn]{revtex4}

\usepackage{amsmath}    % need for subequations
\usepackage{graphicx}   % need for figures
\usepackage{verbatim}   % useful for program listings
\usepackage{color}      % use if color is used in text
\usepackage{subfigure}  % use for side-by-side figures
\usepackage{hyperref}   % use for hypertext links, including those to external documents and URLs
\usepackage{acronym}

\begin{document}

\title{Capture of Inelastic Dark Matter in White Dwarves}

\author{Matthew McCullough}
\email{mccull@thphys.ox.ac.uk}
\affiliation{Rudolf Peierls Centre for Theoretical Physics, University of Oxford, 1 Keble Road, Oxford, OX1 3NP, UK}

\author{Malcolm Fairbairn}
\email{malcolm.fairbairn@kcl.ac.uk}
\affiliation{Physics, King's College London, Strand, WC2R 2LS, UK}

\date{\today}

\begin{abstract}
We consider the capture of inelastic dark matter in white dwarves by inelastic spin-independent scattering on nuclei. We show that if the dark matter annihilates to standard-model particles then, under the assumption of primordial globular cluster formation, the observation of cold white dwarves in the globular cluster M4 appears inconsistent with explanations of the observed DAMA/LIBRA annual modulation signal based on spin-independent inelastic dark matter scattering. Alternatively if the inelastic dark matter scenario were to be confirmed and it was found to annihilate to standard-model particles then this would imply a much lower dark matter density in the core of M4 than would be expected if it were to have formed in a dark matter halo. Finally we argue that cold white dwarves constitute a unique dark matter probe, complementary to other direct and indirect detection searches.
\end{abstract}

\preprint{OUTP-10 28 P}

\maketitle

\section{Introduction}\label{intro}
Many models of dark matter (DM), in particular those models where the DM abundance is a consequence of thermal freeze-out, require the annihilation of DM into standard-model particles which usually implies a non-zero DM-nucleon interaction cross-section.  These two properties can be constrained by a variety of observations.  The annihilation cross-section and products can be constrained by the requirement that DM particles are not overproduced during thermal freeze-out and/or by limits on fluxes of standard model particles that would arise from DM annihilation in the galactic halo.  The DM-nucleon cross-section can be constrained by direct detection experiments which search for the recoil energy deposited when DM scatter off nuclei.  Limits on a combination of the DM-nucleon cross-section and the particular annihilation products can be placed by considering the flux of neutrinos that would result from DM capture and subsequent annihilation in the Sun or Earth.

It has also been pointed out in \cite{Moskalenko:2007ak,Bertone:2007ae} that as white dwarves (hereafter WDs) have no internal energy source, and many cold WDs have now been observed, it is possible to set limits on the DM-nucleon cross-section by showing that the energy released by the annihilation of DM particles in their core could contribute significantly to the luminosity of the star.  %Here we point out one further property of WDs that make them an unique laboratory for DM capture, namely their large escape velocity. 

An interesting possibility for DM-nucleon interactions is that the DM could predominantly scatter off nucleons inelastically to an excited energy state, so-called `inelastic dark matter' (iDM).  This idea has been proposed as a possible explanation \cite{TuckerSmith:2001hy,TuckerSmith:2004jv,Chang:2008gd} for the annual modulation signal observed by the DAMA collaboration \cite{Bernabei:2003za,Bernabei:2008yi}.  The key feature in iDM is the mass splitting, $\delta$, between DM particles that scatter off nuclei.  For iDM explanations of the DAMA/LIBRA annual modulation $\delta$ is of the order $30-130$ keV  (depending on DM mass and couplings).

The main physical consequence of the inelastic splitting is that the minimum velocity for a DM particle to scatter off a nucleus and impart an energy $E_R$ is increased.  This can severely weaken the sensitivity of direct detection experiments as the number of particles in the halo with a large enough velocity to scatter can be very small, or even zero in some cases.  As a consequence the allowed DM-nucleon cross-sections can be orders of magnitude larger than for elastic scattering.

Limits on iDM from capture in the Sun \cite{Nussinov:2009ft,Menon:2009qj,Shu:2010ta} are promising as the escape velocity of the Sun, $v_{esc} \sim 600 - 1300\text{ km s}^{-1}$, is large enough to provide sufficient energy to in-falling DM particles to overcome inelastic splittings $\delta \sim 100$ keV.  Although these limits are hampered by the difficulty of detecting neutrinos, progress in neutrino telescope exposure means strong limits exist on particular models of iDM, for example, limits on sneutrino iDM have been studied in \cite{MarchRussell:2009aq}.  However, models of iDM that annihilate predominantly to $e^+ e^-$, $\mu^+ \mu^-$, $\gamma \gamma$, light hadrons or gluons are immune to these limits as these particles either stop before decaying and producing neutrinos or don't produce neutrinos at all.  In these cases the energy deposited in the Sun from DM annihilations is swamped by the internal energy due to fusion.  Therefore models of the class described in \cite{ArkaniHamed:2008qn} can evade limits from solar capture.

The converse is true for limits on capture in WDs as it is the deposited energy that is used to set the limits, and annihilation to neutrinos will only deposit a small amount of energy in the WD.  Therefore models of DM that attempt to explain the DAMA annual modulation observation with the iDM mechanism, in particular models that also aim to offer an explanation of recent PAMELA \cite{Adriani:2008zr} and Fermi LAT \cite{Abdo:2009zk} observations through DM annihilation, are subject to limits from capture in WDs.

In this paper we investigate limits on the iDM scenario by considering the temperature and luminosity of recently observed WDs in our closest globular cluster M4.  Throughout we assume that the DM annihilates to standard model particles.

\section{White dwarves in M4}\label{WD}
White dwarves are compact objects made up of a degenerate electron core comprised almost entirely of carbon and oxygen.  The electron degeneracy prevents any contraction and the temperature of this core is too low to ignite nuclear fusion reactions.  As a result, WDs have no internal energy source and release only the thermal energy of the non-degenerate ions in the core.  For these reasons the evolution of WDs can be described as a cooling process and the age of a globular cluster such as M4 can be estimated by observing a cut-off at low magnitudes in the WD cooling sequence.  This has motivated observations of WDs in globular clusters down to very low magnitudes.

Recently this low magnitude cut-off has been observed in the globular cluster M4 \cite{Bedin:2009it}.  We use the best-measured set of data from these observations, subject to all of the selection processes detailed in \cite{Bedin:2009it}, including the use of proper motion measurements, to produce a decontaminated sample.

We take the data in the form of magnitudes in the F606W and F775W Hubble Space Telescope filters and convert the colour $m_{606W} - m_{775W}$ to an effective temperature under the assumption that the WD is radiating as a black-body.  To do this we numerically shine a black-body spectrum through the HST filter transmission curves, convert magnitudes to the ACS/WFC Vega-mag system \cite{Bedin:2004zm} and correct for reddening and extinction as detailed in Section 9 of \cite{Bedin:2009it}.  We then use the $m_{606W}$ magnitude, the M4 distance modulus, and the effective temperature of each star and compare with Vega \footnote{We use the parameters $T_{eff}=9550$ K, $R_V = 2.52$ $R_{\odot}$ and $L_V = 37$ $L_\odot$ for Vega.} to calculate the luminosity for each star.    This data is plotted in Fig.\ \ref{Results1}.

Having obtained the luminosity and temperature it is possible to calculate the radius of each WD (under the assumption of black-body radiation).  Then, using the Salpeter equation of state \cite{1961ApJ...134..669S}, we calculate an approximate mass-radius relationship, which can be used to determine an approximate mass for each star.  In Fig.\ \ref{Results2} we show the stars in the mass-luminosity plane.

\section{Dark matter in M4}\label{DM}
In order to set limits on the iDM-nucleon cross-section it is necessary to estimate the DM density surrounding the WD's at the center of M4.  Although it is currently impossible to do this to with any great accuracy, recent developments in the observation and simulated evolution of globular clusters embedded in galactic halos now allow an estimate of the DM content.  Some time ago Peebles \cite{Peebles:1984zz} suggested that globular clusters may be formed in subhalos of DM before falling into galactic halos.  However observations of order $\sim1$ mass-to-light ratios \cite{Meylan:1993yd} and the tidal stripping of stars from some globular clusters \footnote{It should be noted that obvious tidal tails have only been observed in a fraction of the total $\sim 150$ globular clusters residing in our galaxy \cite{Mashchenko:2004hk}.} suggest a significant DM component cannot reside within or without the observed stellar distribution \cite{Moore:1995pb}.  These observations set an upper limit on the DM content of globular clusters.

Recent simulations have shed light on how these results can be reconciled with a primordial scenario of globular cluster formation through the process of tidal stripping.  In fact, the presence of globular clusters has been suggested as a clue towards a resolution of the `missing satellite problem' of cold DM simulations \cite{Cote:2001dj}.  In \cite{Gao:2004au} it was found that once a sub-halo falls into a larger halo mass-loss occurs continually through tidal stripping and the orbit of the sub-halo decays down towards the centre of the larger halo.  Further it was found that the mass-loss can be significant, resulting in only $\sim 2\%$ ($8\%$) of the mass of a sub-halo accreted at $z=2$ ($1$) surviving, and this result appears to be independent of the masses of the halo and subhalo.  The tidal stripping of DM from primordial globular clusters has been investigated with several N-body simulations (see e.g.\ \cite{Mashchenko:2004hk,Moore:2005jj,Saitoh:2005tt,Griffen:2009vg}).  Results suggest that globular clusters can be formed naturally within DM subhalos which are subsequently tidally stripped by the host galaxy, resulting in a baryon-dominated core with a small mass-to-light ratio, resembling observed globular clusters.  In particular a recent analysis of the Aquarius simulation \cite{Griffen:2009vg} lends support to this scenario, and an approximate relation between the current mass of a globular cluster and the mass of the initial subhalo it was embedded within is given as $M_{GC} = 0.0038$ $M_{DM,0}$.

A recent review \cite{Brodie:2006sd} also argues that metal-poor globular clusters formed in low-mass DM halos in the early universe.

The observed cold WDs reside in the dense core of M4, which has survived previous tidal stripping until now.  Therefore it is reasonable to assume that the majority of the DM in the core of M4, well within the tidal radius, will also have survived from the early subhalo.  This assumption is supported by the results of \cite{Mashchenko:2004hk} where it is found that the presence of the stellar core makes the subhalos more resilient to tidal stripping, and for NFW subhalo profiles the DM density in the innermost regions of the subhalo is not modified by the external tidal field.  Outside of the star dominated region the DM subhalo is stripped back to the tidal radius, thus resulting in a mass-to-light ratio close to the purely baryonic value.

Similar reasoning has led to recent consideration of indirect DM signals from DM annihilation in other globular clusters, \cite{PhysRevD.78.027301} and the VERITAS collaboration argue that the association of globular clusters and DM halos fits naturally into the standard paradigm of hierarchical structure formation \cite{Wood:2008hx}.

Motivated by these developments we follow similar methods to those used in \cite{Bertone:2007ae}, to which we refer the reader for details.  The mass of baryonic matter in M4 is estimated to be $M_{b} \sim 10^5 M_\odot$ and the core radius of $0.83'$ in arcminutes implies $r_c = 0.531$ pc when combined with a distance to the cluster of $2.2$ kpc.  The tidal radius is estimated using a concentration parameter of $\log(r_t/r_c) = 1.59$ giving $r_t = 20.66$ pc.  These parameters set the baryon density distribution, which we model with a King profile.

Using cosmological data and taking mass loss during stellar evolution into account, the amount of DM in the original M4 subhalo is estimated to be $M_{DM} \sim 10^7 M_\odot$.  For details of this estimation see \cite{Bertone:2007ae}.  The virial radius, which sets the initial DM distribution is estimated using the fitted form of the spherical collapse overdensity \cite{2006A&A...455...21C};
\begin{equation}
\Delta = \frac{18 \pi^2 + 82 (\Omega_m(z)-1)-39 (\Omega_m(z)-1)^2}{\Omega_m(z)}
\end{equation}
where the matter density is given by \cite{Mashchenko:2004hj}:
\begin{equation}
\Omega_m(z) = \left[1+ \frac{1-\Omega_m}{\Omega_m (1+z)^3}\right]^{-1}
\end{equation}
We take $\Omega_m(0) = 0.273$ giving $\Delta = 357$.  The concentration of low mass halos is given in \cite{Mashchenko:2004hj} as;
\begin{equation}
c(z) = \frac{27}{1+z} \left( \frac{M_{DM}}{10^9 M_\odot}\right)^{-0.08}
\end{equation}
and combining this expression with those for the virial radius, scale radius and central density from \cite{Bertone:2007ae} the original DM subhalos are completely determined by the parameters:
\begin{center}
\begin{tabular}{c|c|c|c}
$z$&$R_{vir} \ [pc]$&$a \ [pc]$&$\rho_c \ [M_\odot \ pc^{-3}]$\\
\hline
0 & 3597 & 92 & 0.37\\
\end{tabular}
\end{center}
We model the original DM halo with an NFW profile \cite{Navarro:1996gj}.  As discussed in \cite{Bertone:2007ae} the core density is a very weak function of the total mass of the subhalo, changing only by a factor $3$ for halo masses between $10^6 M_\odot$ and $10^8 M_\odot$.

It remains to consider the effects of the baryonic core on the DM distribution.  Although the DM density may be enhanced in the core due to the presence of the baryonic core \cite{Blumenthal:1985qy,Gnedin:2004cx,Gustafsson:2006gr} the heating of DM particles due to interactions with stars may tend to wipe out this enhancement.  Therefore by estimating the timescale over which this process occurs with Eqn.\ 3a.\ of \cite{Merritt:2003qk} we can find the radius at which this timescale is equal to the age of the universe.  We find that this radius lies at $r_{heat} = 1.4$ pc and, as this is smaller than the radius where the WDs are observed, we expect heating effects to be small here.  The possible important effect is therefore the contraction of the DM core due to conservation of angular momentum when the gas in the original halo which eventually forms the globular cluster loses energy and falls into the core.  We use the algorithm of Gnedin \cite{Gnedin:2004cx} to perform this baryonic contraction.  Finally as mentioned earlier, to take account of the likely tidal stripping of the stars and DM halo we truncate the density distribution at the tidal radius.

\begin{figure}[h]
\centering
\includegraphics[height=3.2in]{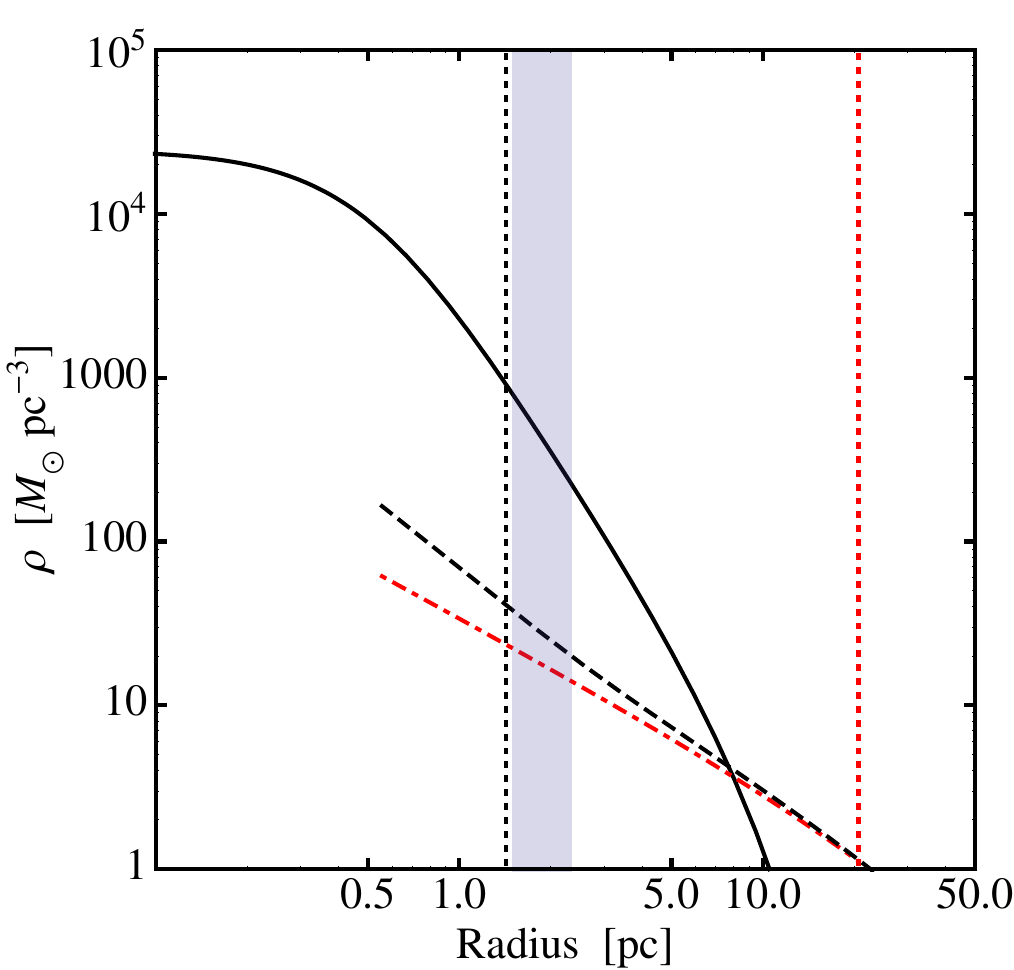}
\caption{Densities of stars (solid line) and the DM halo with and without baryonic contraction effects in dashed black and dot-dashed red respectively.  The region from which the observed WD data is taken is indicated in shaded blue.  The vertical dotted lines denote the radius at which heating effects start to become important (dotted black) and the tidal radius (dotted red).}
\label{densities}
\end{figure}

The estimated DM and star densities for the halos are plotted in Fig.\ \ref{densities}.  One can see that this estimate for the DM density is much smaller than the baryonic density, and in fact for the contracted (uncontracted) halo DM makes up less than $43\%$ ($41\%$) of the total mass of the cluster, consistent with the observed lack of DM in globular clusters.  Further, the total estimated DM content is $7.7 \times 10^4 M_\odot$, less than $1 \%$ of the original $10^7 M_{\odot}$ halo.

We assume a DM density at the largest radius within which the WD data is observed, $r_{max} = 2.3$ pc, giving $\rho_{DM} = 21\text{ M}_\odot \text{ pc}^{-3} = 798 \text{ GeV cm}^{-3}$ for the contracted halo \footnote{The DM density for the uncontracted halo is $\rho_{DM} = 14\text{ M}_\odot \text{ pc}^{-3}$, not much smaller than the contracted halo.}.

\section{Capture of iDM in white dwarves}\label{Capture}
The capture of DM by scattering in stars or planets has been studied for some time, see e.g.\ \cite{1978AJ.....83.1050S,Spergel:1984re,Press:1985ug,Silk:1985ax,Gould:1987ju,Gaisser:1985vn,Gaisser:1986ha,Srednicki:1986vj,Bertone:2007ae}, and recently the capture of iDM in the Sun has been studied \cite{Nussinov:2009ft,Menon:2009qj,Shu:2010ta}.  It is this work which we extend to include capture in WDs and we follow the formalism first set out in \cite{Nussinov:2009ft}, which was subsequently extended to include spin-dependent scattering as well as spin-independent scattering in \cite{Shu:2010ta}.

Recently spin-dependent inelastic scattering has been suggested as a viable alternative to the standard spin-independent iDM scenario \cite{Kopp:2009qt}, where it is shown that spin-dependent couplings to protons, and not neutrons, can give a good fit to the DAMA data whilst remaining consistent with other experiments.  This scenario is subject to limits from scattering in the Sun \cite{Shu:2010ta} however limits from capture in WDs should be weak as WDs are mostly composed of $^{12}$C and $^{16}$O which have no nuclear spin.  It is tempting to consider limits from scattering on $^{13}$C which makes up $\sim 1\%$ of all carbon and has nuclear spin $I=1/2$, however this spin is carried by an unpaired neutron and thus the scenario described in \cite{Kopp:2009qt} will lead to negligible capture rates in WDs.  Therefore we only consider spin-independent scattering iDM in this work.

To calculate the capture rate in a WD we use the equations contained in Section II. of \cite{Shu:2010ta}.  To take account of the incoherent scattering of DM in the nucleus we also use the Helm form factor \footnote{We use the same form factor as in \cite{Shu:2010ta} for consistency, however there is some uncertainty in which form factor is best and care should be taken for processes at high energy transfer \cite{Duda:2006uk}.} and follow \cite{Bertone:2007ae} in making the conservative assumption that the WDs are entirely composed of carbon.  We also assume the DM couplings to neutrons and protons are the same.  This can be corrected for specific models by re-scaling the cross-section accordingly.

To find the DM velocity dispersion we make the assumption of hydrostatic equilibrium and integrate the hydrostatic equation for spherical geometry using the baryon and DM distributions shown in Fig.\ \ref{densities}.  We find that the DM velocity dispersion doesn't exceed $8 \text{ km s}^{-1}$ and, as the capture rate decreases with increasing dispersion, we set $v_0$ to this value.  Similarly the WD velocity through the DM is likely to be of the order of the velocity dispersion and we find this doesn't exceed $6 \text{ km s}^{-1}$, however to set conservative limits we set $v_\star = 20\text{ km s}^{-1}$ which is the escape velocity at the inner radius at which the WDs are observed.

\begin{figure}[]
\centering
\includegraphics[height=3.0in]{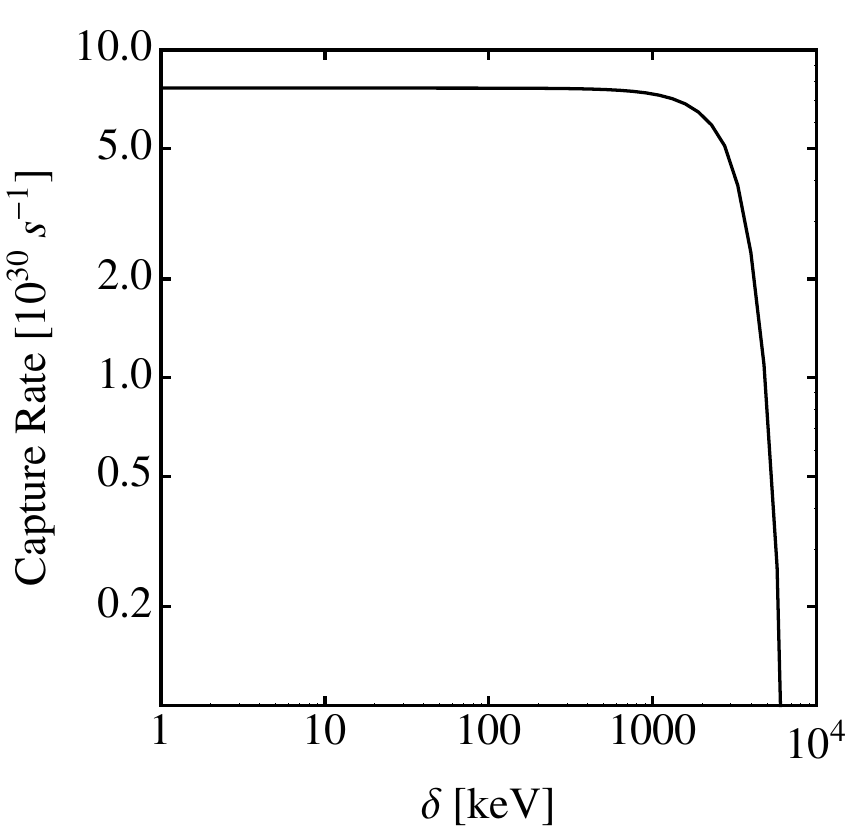}
\caption{The DM capture rate for a solar mass WD as a function of inelastic splitting $\delta$.  The DM parameters are $m_\chi = 50$ GeV and $\sigma_n = 10^{-41}$ cm$^2$.  The capture rate is largely independent of the inelastic splitting up to $\delta \sim 1$ MeV, when it starts to fall off rapidly.}
\label{splittings}
\end{figure}

We calculate the escape velocity and density of nuclei within a given WD using the Salpeter equation of state \cite{1961ApJ...134..669S}.  Due to the large escape velocity of a WD the typical kinetic energy of an in-falling DM particle is of the order $\sim 1$ MeV.  Therefore all of the in-falling particles easily have enough kinetic energy to overcome the inelastic splitting and scatter.  This makes the inelastic splitting relatively unimportant up to splittings $\delta \sim 1$ MeV, where the capture rate starts to decrease.  This is shown in Fig.\ \ref{splittings}.  Although the splittings associated with iDM are much too small to decrease the capture rate significantly we include them in our calculations for the sake of thoroughness.

The capture rate typically falls as the inverse of the DM mass, therefore the luminosity should be largely independent of the DM mass.  However there is a subtle interplay between two factors which leads to a dependence not only on the DM mass but also on the WD mass.  The first factor is due to the conversion between a DM-nucleon and DM-nucleus cross-section at zero momentum transfer, which results in a factor of $\sigma_{\chi N} \propto (\mu_{\chi N}/\mu_{\chi n})^2$ where $\mu$ is the reduced mass and $N$ ($n$) subscripts denote the nucleus (nucleon).  This factor has a preference for heavy DM particles.  However there is also suppression due to the nuclear form factor \footnote{Due to the form factor suppression energy transfers of $E_R > 4$ MeV contribute very little to the capture rate, even if kinematically possible.}.  Therefore, although heavier DM particles have a larger range of scattering energies, the higher energy events are suppressed.  This effect therefore discriminates against heavy DM particles.

\begin{figure}[]
\centering
\includegraphics[height=3.0in]{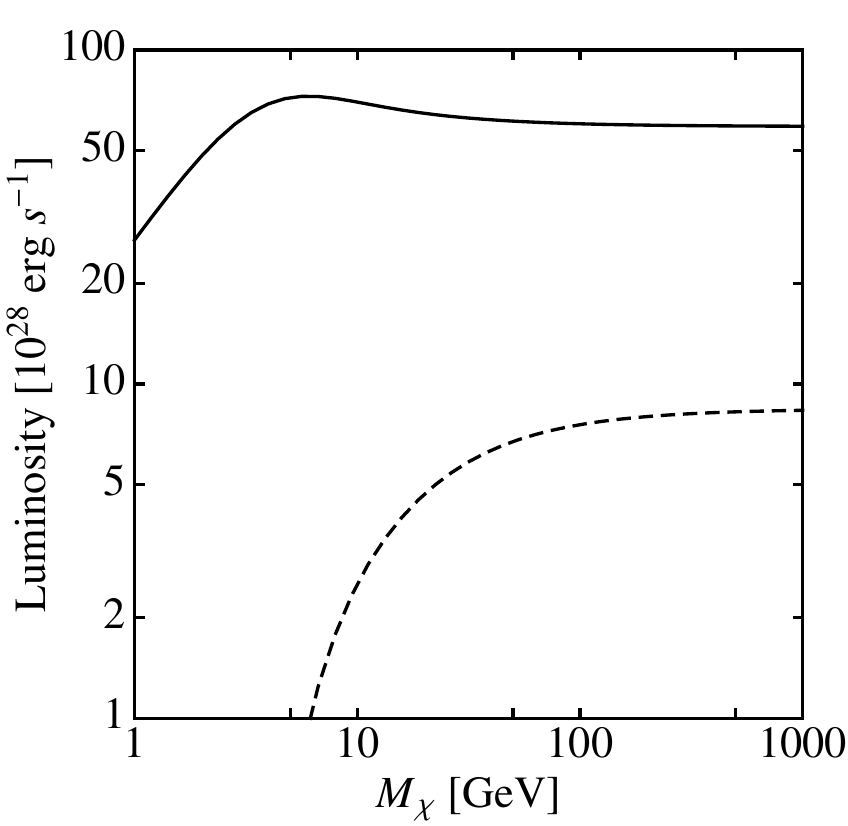}
\caption{The luminosity of a WD due to capture and annihilation of DM as a function of DM mass for WDs of mass $M_\star = M_\odot$ (solid line) and $M_\star = 0.2$ $M_\odot$ (dashed).  The DM parameters are $\delta = 130$ keV and $\sigma_n = 10^{-41}$ cm$^2$.  There is a slight enhancement around $m_\chi \sim 10$ GeV for the solar mass WD and the luminosity is suppressed for small DM masses.  This behavior is discussed in the text.}
\label{masses}
\end{figure}

Which of these two factors wins out depends on the WD mass.  As heavy WDs have greater escape velocities ($\sim (7 - 12)\times 10^3$ km s$^{-1}$), heavy DM particles feel the form factor suppression more and light DM particles lead to greater luminosities.  For light WDs the escape velocities are lower ($\sim (2 - 3)\times 10^3$ km s$^{-1}$) the form factor suppression is subdominant and heavier DM particles lead to a greater luminosity.  The mass dependence for two different WDs is illustrated in Fig.\ \ref{masses}.

As described in \cite{Bertone:2007ae} the time-scale for equilibrium between the capture and annihilation of DM in WDs is roughly of the order of one year, and therefore we can safely assume that the capture rate is one-half of the annihilation rate.  We also assume that all of the energy of the annihilating particles contributes to the black-body luminosity of the WDs, however specific DM models could be investigated by calculating the fraction of energy lost as neutrinos per annihilation and weakening the limits accordingly.

%% Relevant equations from the Menon et al paper are in the comment below.
\begin{comment}
For an inelastic splitting of $\delta$ the differential capture rate for a volume element of a star is given by
\begin{equation}
\frac{dC_\odot}{dV} = \int^{\infty}_0 f(u) du \frac{w}{u} \Omega^-_v(w) \Theta(\frac{1}{2} m_\chi w^2-\delta \frac{m_\chi+m_N}{m_N})
\end{equation}
where
\begin{equation}
f(u) du = \frac{4 \pi n_{DM} \tilde{v}^2}{N(\infty) u v_{\star}} \exp(-\frac{u^2+v^2_{\star}}{\tilde{v}^2}) \sinh(\frac{2 u v_{\star}}{\tilde{v}^2}) du
\end{equation}
and $v_\star$ is the velocity of the star w.r.t.\ the DM rest frame, $\tilde{v}$ is the velocity dispersion of the DM, $w^2=u^2+v^2(r)$ with $v(r)$ the escape velocity inside the star at radius $r$ and $N(\infty)$ is the velocity distribution normalization factor.

We also have
\begin{equation}
\Omega^-_v (w)= \frac{\sigma n w}{E^{elastic}_{max}} \int^{\Delta E_{max}}_{\Delta E_{min}} d(\Delta E) F^2(\Delta E) \Theta (\Delta E -E_\infty)
\end{equation}
where $F(\Delta E)$ is the nuclear form factor, for which we use the Helm form factor from \cite{Lewin:1995rx}, $E = m_\chi w^2/2$, $E_\infty = m_\chi u^2/2$ and $E^{elastic}_{max} = 2 \mu^2 v^2 /M_N$.  Expressions for $\Delta E_{max}$, $\Delta E_{min}$ and $\sigma$ can be found in \cite{Menon:2009qj} and also depend on $\delta$.
\end{comment}

\section{Uncertainties}\label{Uncertainties}
We will now discuss some of the assumptions that have gone into this calculation.  The greatest source of uncertainty is the estimation of the DM density.  It should be noted that there are some models where globular clusters are not formed due to the collapse of a DM dominated halo, in particular the observations made by Gilmore and collaborators have lead them to argue that low mass star clusters are fundamentally different to higher mass galaxies rather than both being members of a continuous family \cite{Gilmore:2007fy}.  The explanation for this scenario typically requires some kind of modification of DM such as warm DM, a cold/hot admixture or a non-zero self-interaction cross section such that there is a minimum size for dark matter halos in the Universe.  Since we are trying to put constraints on models of cold DM, it is a consistent assumption that the globular clusters do form in the centre of DM halos in the early Universe but we note that this is an uncertainty.

It has been shown that for direct detection experiments the details of the DM velocity distribution can have a significant impact on detection rates \cite{Fairbairn:2008gz}, particularly for iDM \cite{MarchRussell:2008dy,Kuhlen:2009vh}.  However, due to the large escape velocity of the WDs all in-falling DM particles will have a large enough velocity to scatter and the details of the velocity distribution will be unimportant.

We have made the estimate that the WDs are traveling at the local escape velocity, however a more realistic (but less conservative) assumption would be that they are traveling at a speed closer to the local velocity dispersion, which is roughly a factor of 3 smaller.  As the capture rate is inversely proportional to this speed we may have underestimated the capture rate by the same amount. 

The observed WDs may not be entirely composed of Carbon, and may contain heavier elements, however this assumption is safe as the capture rate is smallest for light target nuclei.  Due to the large escape velocity the energy transfer in scattering events can be large ($\sim 1$ MeV), and the scattering is thus significantly suppressed by the nuclear form factor.  We use the Helm form factor, which for light nuclei can be up to $\sim 20 \%$ greater than more realistic form factors at these high energy transfers \cite{Duda:2006uk}.  Therefore a conservative estimate of the uncertainty due to the choice of form factor is of the order $20 \%$.

For considerations relating to errors in WD observations we refer the reader to \cite{Bedin:2009it}.

\section{Results}\label{Results}
\begin{figure}[]
\centering
\includegraphics[height=3.0in]{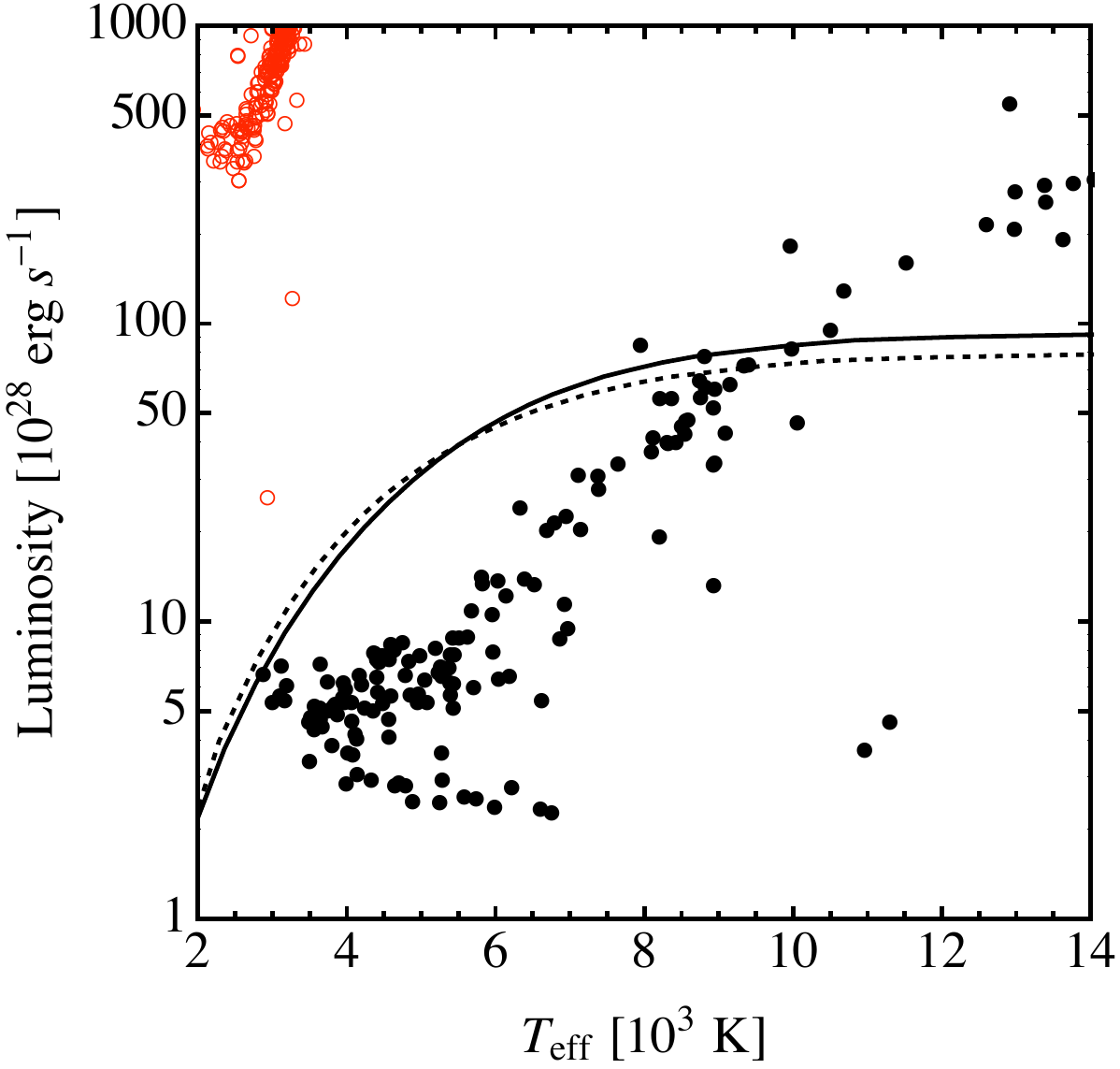}
\caption{Observed WDs (black dots) and main-sequance stars (open red circles).  Also plotted are the luminosity and temperature of WDs in the mass range $0.1-1.35$ $M_\odot$ for the two benchmark points i1 and i2 (dashed) described in the text.  For both curves a DM-nucleon cross-section of $10^{-41}$ cm$^2$ is assumed.  The luminosity from DM capture is greater than a significant number of the observed WD luminosities.}
\label{Results1}
\end{figure}

In Fig.\ \ref{Results1}. we show the observed WDs in the Temperature-Luminosity plane.  On the same plot we show curves for WDs whose sole energy source is due to DM annihilation in the core for WDs ranging in mass from $0.1$ $M_\odot$ to $1.35$ $M_\odot$.  These curves correspond to two benchmark points:
\begin{center}
\begin{tabular}{c|ccc}
 & $M_\chi$ [GeV] & $\delta$ [keV] & $\sigma_n$ [$10^{-41}$ cm$^{-1}$] \\
\hline
i1 & 10 & 40 & $1$ \\
i2 & 100 & 130 & $1$ \\
\end{tabular}
\end{center}

The mass and splitting for the point i1 corresponds roughly to best-fit points for channeled scattering on Iodine and in \cite{SchmidtHoberg:2009gn,Kopp:2009qt} it is shown that for these parameters consistency with other direct detection experiments is possible.  The mass and splitting in point i2 corresponds to the conventional quenched scattering iDM scenario.  This scenario can be considered ruled out by other direct detection experiments, however this depends sensitively on how (or whether) data from different experimental runs is combined, the inclusion of borderline scattering events in the CRESST detector, choice of Iodine quenching factor and details of the DM velocity distribution \cite{SchmidtHoberg:2009gn,MarchRussell:2008dy,Kuhlen:2009vh}.  Therefore for completeness we still include this choice of parameters in our analysis.  We choose a cross-section of $\sigma_n = 10^{-41}$ cm$^2$ as this is below the cross-section at which the optically thick limit applies and the capture rate becomes independent of the scattering cross-section.  This important point was first emphasized in the context of inelastic dark matter capture in \cite{Hooper:2010es} which appeared shortly after the initial preprint of the current work was placed on the ArXiv. We have updated our cross-sections so as to stay below this limit although the main conclusions of this paper are unchanged.

\begin{figure}[h]
\centering
\includegraphics[height=3.0in]{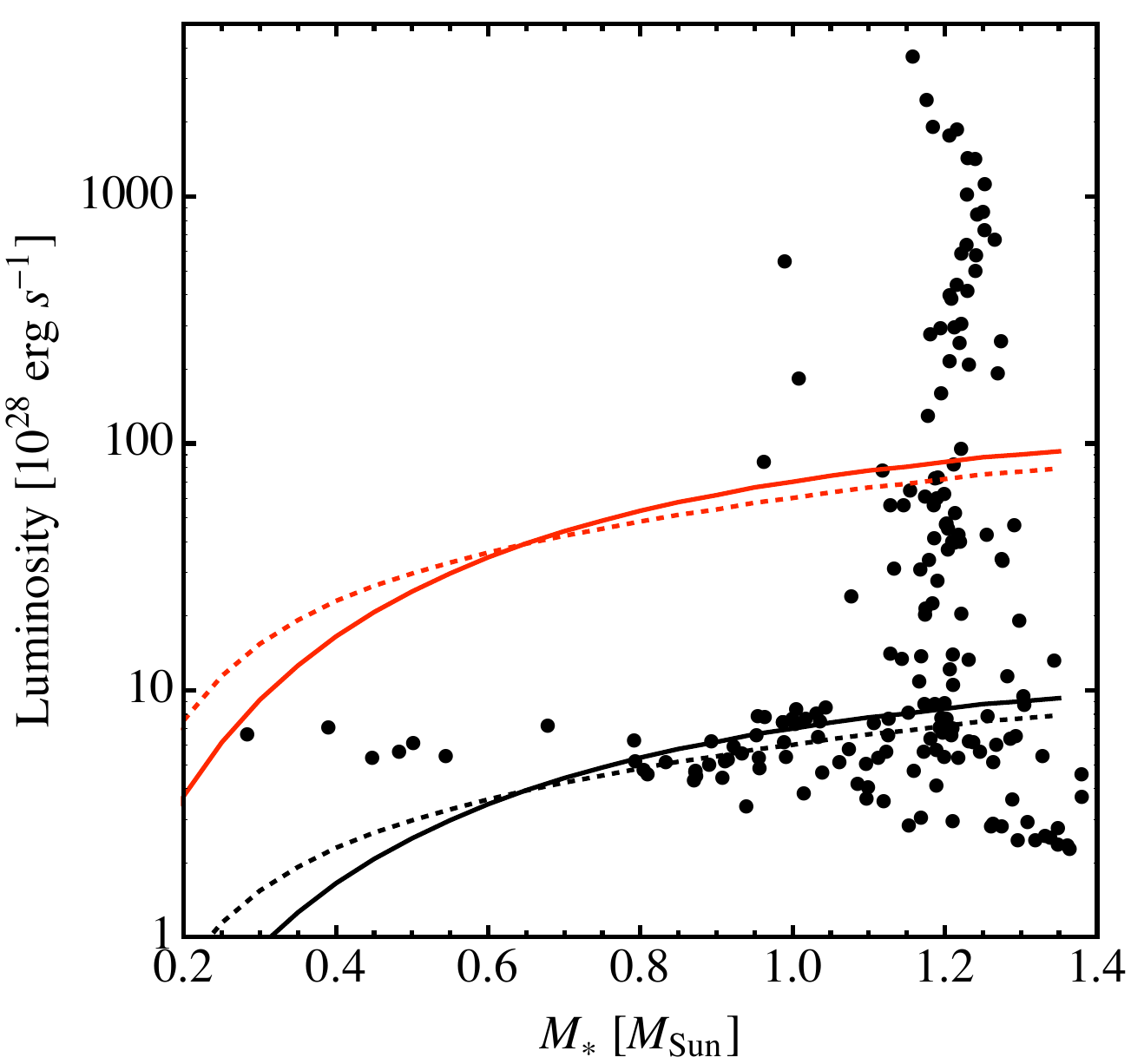}
\caption{Observed WDs (black dots).  Also plotted are the luminosity and temperature of WDs in the mass range $0.1-1.35$ $M_\odot$ for the two benchmark points i1 and i2 (dashed).  The upper red curves correspond to a cross-section of $\sigma_n = 10^{-41}$ cm$^2$ and the lower black curves to $\sigma_n = 10^{-42}$ cm$^2$.}
\label{Results2}
\end{figure}

As the temperature of a black-body is related to the luminosity as $T\propto L^{1/4}$ Fig.\ \ref{Results1} can be misleading, as a change in cross-section does not correspond to a simple re-scaling of the luminosity.  Therefore in Fig.\ \ref{Results2} we estimate the observed WD masses as described in Section.\ \ref{WD}, and plot curves showing the luminosity due to DM capture and annihilation for a given WD mass.

As can be seen the low luminosities and temperatures of the observed WDs appear incompatible with cross sections greater than $\sigma_n \sim 10^{-42}$ cm$^2$ for either of the benchmark points.  A cross-section below $10^{-42}$ cm$^2$ could possibly be argued as acceptable given the uncertainties, and a cross-section as low as $10^{-43}$ cm$^2$ would evade these bounds entirely.  However any cross-section greater than $10^{-42}$ cm$^2$ would certainly appear to contradict the existence of such cold WDs in M4.

Recent analyses of iDM find a best fit cross-section of $2\times10^{-37}$ cm$^2$ \cite{Kopp:2009qt} and $4.7\times10^{-39}$ cm$^2$ \cite{SchmidtHoberg:2009gn} for the benchmark point i1.  As these cross-sections are greater than $\sim10^{-42}$ cm$^2$ both scenarios appear excluded by the analysis above, and even the lowest cross-section found for the channeled iodine region of $9\times10^{-42}$ cm$^2$ \cite{SchmidtHoberg:2009gn} would be difficult to reconcile with these results.  The conventional iDM scenario of quenched scattering off iodine (benchmark point i2) typically requires cross-sections greater than $\sim 3\times10^{-40}$ cm$^2$, with a best-fit point at $\sigma_n = 10^{-38}$ cm$^2$ \cite{Kopp:2009qt}.  This scenario seems excluded by this analysis.

These constraints can be evaded in any combination of the following scenarios:
\begin{itemize}
\item  The DM density is less than $\sim 1\%$ of that estimated here.  This appears plausible, however it would imply that DM makes up less than $\sim 1\%$ of the total mass of the globular cluster.  If globular clusters are born in dark matter halos it is hard to imagine how the ratio of DM to baryonic matter in M4 could be so far below cosmological values.  If M4 did not form in a DM halo it is likely the DM density would be low enough to evade these bounds.
\item  The DM could annihilate predominantly to neutrinos, thus contributing little to the visible luminosity of the WDs.  This scenario would be difficult to reconcile with limits on neutrino fluxes from DM annihilation in the Sun \cite{Nussinov:2009ft,Menon:2009qj,Shu:2010ta}.
\item  The iDM couples to nuclei only through spin-dependent interactions as recently suggested \cite{Kopp:2009qt}.  If iDM couples to neutrons then it may be possible to set limits by considering scattering off $^{13}$C, however as this scenario already appears disfavoured by direct detection experiments \cite{Kopp:2009qt} this is not investigated here.  If the coupling is only to protons then limits from WDs pose little threat.
\end{itemize}

From simple assumptions about the DM density in globular clusters (including their formation in DM halos), the composition of cold WDs, and the capture of iDM we argue that the iDM explanation of the annual modulation observed by the DAMA collaboration is incompatible with the observation of cold WDs in M4 if the DM annihilates to standard model particles.  Alternatively if the inelastic dark matter scenario were to be confirmed and it was found to annihilate to standard-model particles then this would imply a much lower dark matter density in the core of M4 than would be expected if it were to have formed in a dark matter halo.

\section{Discussion}
We now discuss some of the salient features of white dwarves which make them a unique probe of DM.

It is interesting to note that in the case of elastic DM scattering cross-sections $\sigma_n \sim 10^{-43}$ cm$^2$ evade the WD bounds \footnote{The difference between the conclusions presented here and those in \cite{Bertone:2007ae} arises mostly due to the form factor suppression which has been included in this analysis.}.  It is unlikely that observations of WDs much cooler than those in M4 will be made as the low luminosity cut-off has been observed, and the luminosity of WDs is limited by the age of the Universe.  Therefore it is unlikely that limits from WDs will ever compete with direct detection limits for weak-scale elastic DM.

However WDs constitute unique DM probes for three reasons:
\begin{itemize}
\item  The large escape velocity enables in-falling DM particles to easily overcome inelastic splittings and leads to large energy transfers in scattering.
\item  The low mass of carbon gives WDs sensitivity to light DM scenarios, where most direct detection experiments lose sensitivity.
\item  Limits from capture in the Sun arise due to neutrino annihilation products and are therefore insensitive to DM annihilating to $e^+ e^-$, $\mu^+ \mu^-$, $\gamma \gamma$, light hadrons or gluons.  It is specifically this scenario where limits from WDs are strongest.
\end{itemize}

We have only considered iDM capture in this work, however numerous possibilities exist for future study of DM capture in WDs.  Examples would include DM with mass splittings of the order a few MeV or DM which scatters through a light mediator, $m_\phi \sim$ MeV, which could be enhanced in WDs through the propagator $1/(q^2-m_\phi^2)$.

Finally we note that if DM were to be discovered in future experiments, and details of the DM-nucleon cross-section and annihilation products were to be established, then cold WDs could be used to determine an upper limit on the DM density within M4, thus giving clues as to the formation of globular clusters.

\section{Acknowledgements}
We thank Andy Eyre and Brad Hansen for conversations and John March-Russell for suggesting the investigation of iDM capture in astrophysical objects.  We are also grateful to Maurizio Salaris for conversations and providing the WD data.  MM is supported by an STFC Postgraduate Studentship and both MF and MM acknowledge support from the EU Marie Curie Network ÒUniverseNetÓ (HPRN-CT-2006-035863).

\bibliography{iDMWDrefs}

\end{document}